\documentclass[prb,showpacs,byrevtex,floatfix,twocolumn,superscriptaddress]{revtex4}
\usepackage{amssymb,graphicx,afterpage}
\begin{document}
\title{\boldmath Optical phase diagram of perovskite-type colossal magnetoresistance manganites\\ with near-half doping \unboldmath}
%
%
\author{I. K\'ezsm\'arki}
\affiliation{Department of Applied Physics, University of Tokyo,
Tokyo 113-8656, Japan} \affiliation{Spin Superstructure Project,
ERATO, Japan Science and Technology Agency (JST), Tsukuba 305-8562,
Japan} \affiliation{Department of Physics, Budapest University of
Technology and Economics, 1111 Budapest, Hungary}
\author{Y. Tomioka}
\affiliation{Correlated Electron Research Center (CERC), National
Institute of Advanced Industrial Science and Technology (AIST),
Tsukuba 305-8562}
\author{S. Miyasaka}
\affiliation{Department of Applied Physics, University of Tokyo,
Tokyo 113-8656, Japan}
\author{L. Demk\'o}
\affiliation{Department of Physics, Budapest University of
Technology and Economics, 1111 Budapest, Hungary}
\author{Y. Okimoto}
\affiliation{Correlated Electron Research Center (CERC), National
Institute of Advanced Industrial Science and Technology (AIST),
Tsukuba 305-8562} \affiliation{Department of Materials Science,
Tokyo Institute of Technology, Meguro-ku, Tokyo 152-8551, Japan}
\author{Y. Tokura}
\affiliation{Department of Applied Physics, University of Tokyo,
Tokyo 113-8656, Japan} \affiliation{Spin Superstructure Project,
ERATO, Japan Science and Technology Agency (JST), Tsukuba 305-8562,
Japan} \affiliation{Correlated Electron Research Center (CERC),
National Institute of Advanced Industrial Science and Technology
(AIST), Tsukuba 305-8562}
\date{\today}
%
%
\pacs{\ }
\begin{abstract}
We present a systematic optical study for a bandwidth-controlled
series of nearly half doped colossal magnetoresistive manganites
RE$_{0.55}$AE$_{0.45}$MnO$_3$ (RE and AE being rare earth and
alkaline earth ions, respectively) under the presence of quenched
disorder over a broad temperature region $T=10-800$\,K. The ground
state of the compounds ranges from the charge and orbital ordered
insulator through the spin glass to the ferromagnetic metal. The
enhanced phase fluctuations, namely the short-range charge and
orbital correlations dominate the paramagnetic region of the phase
diagram above all the ground-state phases. This paramagnetic region
is characterized by a full-gap to pseudo-gap crossover towards
elevated temperatures where a broad low-energy electronic structure
appears in the conductivity spectra over a large variation of the
bandwidth. This pseudo-gap state with local correlations is robust
against thermal fluctuations at least up to $T=800$\,K. For small
bandwidth the onset of the long-range charge order is accompanied by
an instantaneous increase of the gap. The emergence of the
ferromagnetic state is manifested in the optical spectra as a
first-order insulator to metal transition for compounds with
moderate bandwidth while it becomes a second-order transition on the
larger bandwidth side. Unusually large scattering rate of the
metallic carriers is observed in the ferromagnetic state which is
attributed to orbital correlation with probably rod-like
($3z^2-r^2$-like) character.
\end{abstract}
\maketitle
%
%
\section{Introduction}

The accumulation of dopant charges in Mott insulators and the
dependence of the critical doping level on the relative strength of
the Coulomb repulsion have been studied for a large variety of
correlated electron
systems.\cite{Katsufuji95,Okimoto95,Katsufuji97,Tokura98} Generally,
with increasing $U/W$ ratio (where $U$ is the on-site Coulomb
interaction and $W$ the effective one-electron bandwidth) the
insulating phase becomes more robust against doping-induced
insulator to metal transition.

In perovskite-type manganites, due to the strong interplay between
electronic charge, spin, orbital, and lattice degrees of freedom, at
specific carrier concentrations (so-called commensurate doping
levels) several charge and orbital ordered (CO/OO) patterns emerge
accompanied by antiferromagnetic
ordering.\cite{Goodenough55,Jirak85} These CO/OO states are
inherently insulators and in their vicinity charge and orbital
correlation effects are critically intensified. Consequently, for
the compounds with relatively narrow one-electron bandwidth the
CO/OO ground state remains stable against considerable variation of
doping and hence the metallic conduction does not
appear.\cite{Okimoto00,Tomioka02,Tomioka04} However, at a fixed
carrier concentration, as the bandwidth of the $e_g$ electrons is
increased, the CO/OO suddenly collapses and the compound goes
through a first order insulator to metal transition and
simultaneously an antiferromagnetic to ferromagnetic
transition.\cite{Tomioka02,Tomioka04} This tendency is discerned in
Fig.~1(a) in the bandwidth-temperature phase diagram typical of
RE$_{1-x}$AE$_x$MnO$_3$ materials close to half doping, namely for
$x=0.45$ in the present case. Throughout the paper we shall refer
the averaged radius of RE$_{1-x}$AE$_x$ as a measure of the $e_g$
one-electron bandwidth, in the sense that it reflects the magnitude
of the double exchange interaction responsible for the ferromagnetic
metallic phase (FM).\cite{Furukawa95} The competition of
spin-orbital exchange interaction energies, further balanced by
strong coupling to the lattice, is clearly manifested in the
first-order nature of the transition between the two robust phases
with ordering temperatures ($T_{CO}$ and $T_c$, respectively) higher
than $T=200$\,K. The first-order transition line separating the two
ordered phases is terminated by a bicritical point at as high
temperature as $T_{bicr}\approx200$\,K, indicating the stability of
the both phases against thermal fluctuations. In contrast, the
application of magnetic fields of a few tesla efficiently induce the
CO/OO AFI to FM transition implying the precise balance of the free
energy for the two neighboring ground
states.\cite{Tomioka96,Tokura00,Okimoto00,Tomioka02,Tomioka06} These
result in gigantic phase fluctuations above the bicritical point and
cause that the onset of the ordered phase with decreasing
temperature also occurs in a first-order-transition manner on either
side of the bicritical
point.\cite{Murakami03,Adams04,Kim02,Tomioka04} Among macroscopic
quantities, colossal magnetoresistance (CMR) is the most dramatic
manifestation of this phase-fluctuation induced
phenomenon.\cite{Tokura-book,Tokura06}
\begin{figure}[h!]
\includegraphics[width=2.6in]{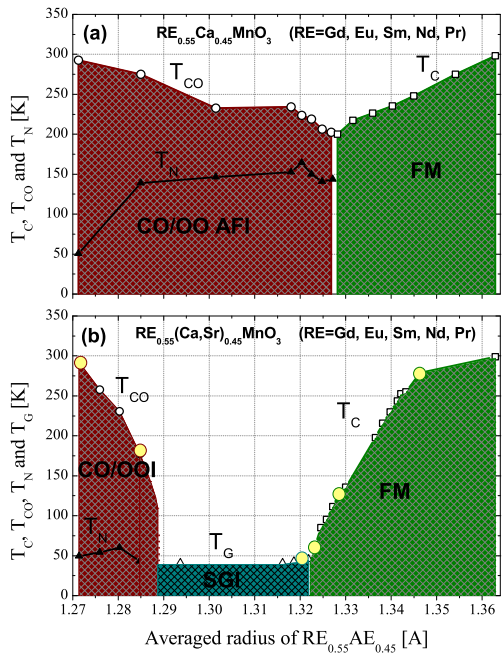}
\caption{(Color online) Bandwidth--temperature phase diagram of
colossal magnetoresistance manganites near half doping with low and
high level of quenched disorder displayed in the upper and lower
panel, respectively (reproduced from Tomioka et al.
\cite{Tomioka04}). In the former, $T_{CO}$ and $T_C$ meet at
$\sim200$\,K and form the so-called bicritical point, while in the
latter the ordered phases are suppressed and an intermediate spin
glass phase appears below $T_G$ due to the quenched disorder. The
compounds investigated in the present optical study are highlighted
in the lower panel.} \label{fig1}
\end{figure}

In perovskite-type manganites the carriers are doped via the partial
substitution of the rare earth component (RE) with divalent ions in
the form of RE$_{1-x}$AE$_x$MnO$_3$ where usually AE$=$Ca, Sr, and
Ba. In addition to the change of the band filling, this random
doping introduces quenched disorder into the lattice which can
result in a critical suppression of the long range order on the both
sides of the bicritical point (Fig. 1(b)). The variance of the ionic
radius at the perovskite $A$ site was found to well represent the
degree of disorder.\cite{Rodriguez98,Rodriguez00} For the compounds
in Fig.~1(a) the variance is rather low while it is considerably
large for those in Fig.~1(b) and therefore they will be referred to
as low- and high-disorder phase diagram, respectively.

The long-range CO/OO AFI state is primarily sensitive to the local
lattice distortion and looses territory towards larger variance. The
broad intermediate region bounded by the two ordered phases in
Fig.~1(b) is characterized by short-range CO/OO correlation enhanced
toward low temperatures and finally frozen to an insulating glassy
state below $T_G$. The weakening of the ferromagnetism or the
critical suppression of $T_c$ close to the bicritical point is
probably the result of CO/OO fluctuations, as well. This was indeed
observed in recent x-ray diffuse scattering and Raman scattering
experiments: Similarly to the temperature region above $T_{CO}$ and
$T_G$,\cite{Shimomura99,Shimomura00,Jirak00,Mathieu04} for larger
bandwidth the CO/OO correlation is also critically enhanced toward
low temperatures where it was suddenly cut by the first-order FM
transition.\cite{Tomioka03,Motome03} Therefore, it seems very
plausible to consider that the paramagnetic region above all the
ground-state phases is essentially governed by strong short-range
correlation of the CO/OO phase in addition to FM spin fluctuations.

In an effort to understand the details of the CMR effect for nearly
half-doped perovskite manganites with high quenched disorder --
beyond the basic picture capturing the role of competing magnetic
exchange interactions and interlocking of spin, charge, orbital, and
lattice degrees of freedom \cite{Furukawa94,Millis95} -- several
intrinsic and extrinsic mechanisms have been proposed. The formers
treat CMR as a result of strong fluctuations of a homogeneous
quantum phase,\cite{Tokura00,Murakami03,Mathieu04} while the latters
consider disoder-induced phase separation and percolation effects
indispensable for the description of the
phenomenon.\cite{Uehara99,Dagotto02} Recently, the existence of
topological defects in the orbital order, namely orbital solitons
carrying $\pm e/2$ charge, for manganites close to half doping were
also suggested as a natural explanation of nanoscale inhomogeneities
in these materials.\cite{Brey05}
\begin{figure}[h!]
\includegraphics[width=2.6in]{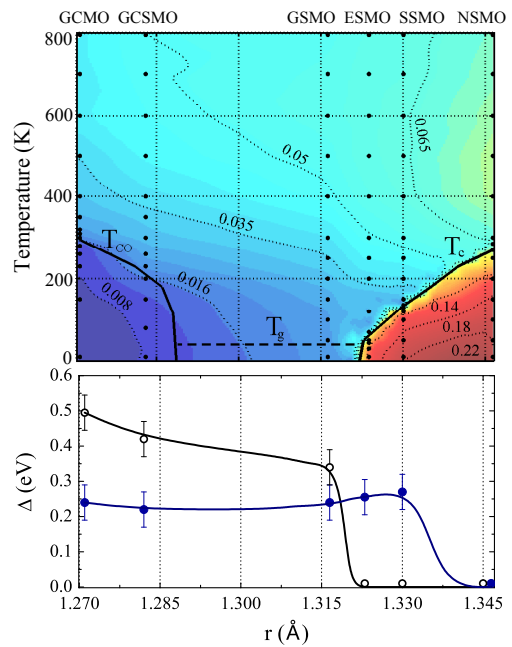}
\caption{(Color) Optical phase diagram of nearly half-doped
RE$_{0.55}$AE$_{0.45}$MnO$_3$ CMR manganites. Upper panel: contour
map for the low-energy spectral weight, i.e.
$N_{eff}(\omega=0.68eV)$, over the average ionic radius ($r$) vs
temperature plane. The map is obtained by the interpolation of the
experimental data indicated by black dots. Besides the color
scaling, numerical values are also given on the contour lines for
reference. Phase boundaries are also indicated by showing the CO/OO,
SG, and FM transition temperatures. Lower panel: variation of the
charge gap $\Delta$ with the ionic radius (equivalent to the
effective one-electron bandwidth). Full symbols represent the gap
just above $T_{CO}$, $T_c$ and at $T=200$\,K for GSMO. The
ground-state ($T=10$\,K) gap is also plotted by open cicrles. The
solid lines are merely the guide to the eyes.} \label{phasediag}
\end{figure}

In order to investigate the influence of the dramatic phase change
and enhanced CO/OO fluctuations on the low-energy electronic states
of CMR manganites, we performed a systematic optical study over the
compounds highlighted in Fig.~1b. Changing the chemical composition
along this series results dominantly in a variation of the bandwidth
while the level of quenched disorder is kept relatively high. The
presence of neighboring phases with different types of orbital order
in nearly half-doped perovskite manganites
\cite{Konishi99,Tobe04,Endoh05,Tokura06} also gives the hint that
critical amplification of orbital fluctuations may play a unique
role in CMR effect. The main goal of the paper is to follow the
fingerprints of short-range CO/OO correlation in the low-energy
optical conductivity spectra of RE$_{0.55}$AE$_{0.45}$MnO$_3$
compounds over a broad range of the bandwidth-temperature phase
diagram. The systematics of the low-energy electronic structure,
obtained in the course of the present optical study and discussed in
the body of the paper, are visualized in Fig.~\ref{phasediag}.

\section{Experimental details}
\label{exp}

All of the RE$_{0.55}$AE$_{0.45}$MnO$_3$ samples (RE$=$Nd, Sm, Eu,
and Gd; AE$=$Ca and Sr) investigated here were single crystals grown
by a floating-zone method. Throughout the paper, for
Nd$_{0.55}$Sr$_{0.45}$MnO$_3$, Sm$_{0.55}$Sr$_{0.45}$MnO$_3$,
Eu$_{0.55}$Sr$_{0.45}$MnO$_3$, Gd$_{0.55}$Sr$_{0.45}$MnO$_3$,
Gd$_{0.55}$(Ca$_{0.75}$Sr$_{0.25}$)$_{0.45}$MnO$_3$,
Gd$_{0.55}$Ca$_{0.45}$MnO$_3$ we use the abbreviations NSMO, SSMO,
ESMO, GSMO, GCSMO, and GCMO, respectively. In the CO/OO state, these
compounds are characterized by electron dynamics with nearly
isotropic optical conductivity within the $ab$-plane and with
reduced low-energy spectral weight for polarization along the
$c$-axis.\cite{Tobe04} Therefore, in the present systematic study of
the optical properties over a broad range of the
bandwidth-temperature plane, we used oriented single crystals and
measured reflectivity spectra at nearly normal incidence on the
(001) crystallographic plane (in the pseudocubic setting). The
samples were cut and polished with alumina powder to the optical
flatness. In order to eliminate residual strains of surface layers
induced by mechanical polishing the crystals were annealed at
$1300^{\circ}$C for about $30$ hours.\cite{Saitoh99}

Reflectivity spectra were investigated in a photon-energy range of
$E=3$\,meV$-$$6$\,eV below room temperature and in a slightly
limited range ($E\geq80$\,meV) for $T=10$$-$$800$\,K. For the proper
Kramers-Kronig analysis the room temperature measurements were
extended up to $40$\,eV with use of synchrotron radiation at UV-SOR,
Institute for Molecular Science. Although the obtained conductivity
spectra cover the whole range of representative charge-transfer
excitations,\cite{Arima95} here we focus on the low-energy part
governed by correlation effects. In our convention $\sigma(\omega)$
denotes the real part of the optical conductivity simply referred to
as the optical conductivity.

\section{Evolution of the ground state}
\label{ground_state}

\begin{figure}[th]
\includegraphics[width=2.65in]{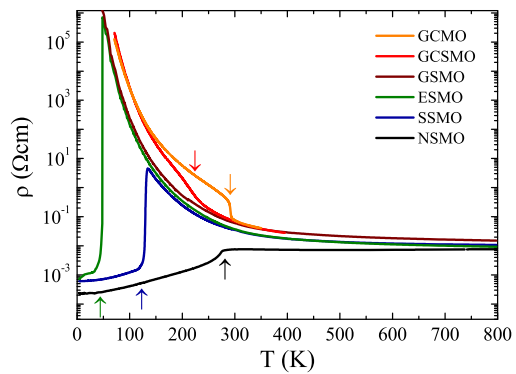}
\caption{(Color online) Temperature dependence of the resistivity
for RE$_{0.55}$AE$_{0.45}$MnO$_3$ single crystals (RE$=$Nd, Sm, Eu,
and Gd; AE$=$Ca and Sr). The arrows indicate the first-order
transition to the ferromagnetic ground state for NSMO, SSMO, and
ESMO and the onset of the charge ordered state in GCSMO and GCMO.}
\label{fig2}
\end{figure}

The temperature dependence of the dc resistivity measured on the
respective single crystals by the standard four-probe method is
shown in Fig.~\ref{fig2}. The onset of the CO/OO or the FM
transition is clearly manifested in the resistivity curves as
indicated by arrows. As a general tendency, with increasing
bandwidth $T_{CO}$ steeply decreases. In case of
Gd$_{0.55}$Sr$_{0.45}$MnO$_3$ with the glassy ground state, although
the long-range CO/OO order is lost, the system remains insulating.
Upon further increase of the bandwidth, the low-temperature phase
becomes a ferromagnetic metal as discerned in Fig.~\ref{fig2} for
R$=$Eu, Sm, and Nd. Irrespective of the ground-state nature, the
$\rho(T)$ curves tend to converge in the high-temperature disordered
phase. In fact, the difference at room temperature is only one order
of magnitude and it is further reduced down to a factor of $\sim2$
at the elevated temperature, $T=800K$.

\begin{figure}[th!]
\includegraphics[width=2.6in]{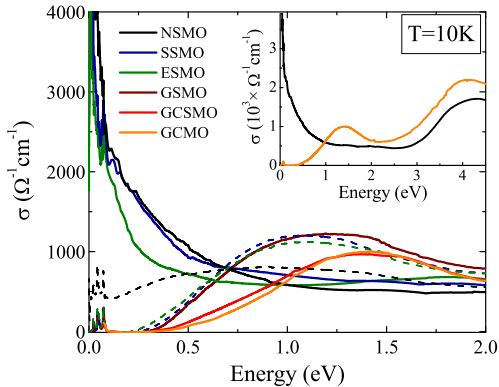}
\caption{(Color) Comparison of the ground state optical conductivity
spectra for a bandwidth-tuned series of
RE$_{0.55}$AE$_{0.45}$MnO$_3$ manganites (full lines). The dashed
lines correspond to optical conductivity of NSMO, SSMO, and ESMO in
their most insulating state, i.e. just above $T_c$. The inset shows
the conductivity spectra of the two end-compounds on a wider energy
scale.} \label{fig3}
\end{figure}

For the study of the bandwidth-controlled low-energy spectral
changes, we mostly focus on the intraband transitions of the Mn $3d$
electrons (hybridized with O $2p$ states) located below $\sim2$\,eV.
The ground state conductivity spectra for the two end-compounds are
plotted over a broader energy range up to $\sim4.5$\,eV in the inset
of Fig.~\ref{fig3}. Gd$_{0.55}$Ca$_{0.45}$MnO$_3$ is an insulator
with a charge gap $\Delta_{cg}\approx0.5$\,eV. The broad peak
located at around $E_{peak}=1.4$\,eV corresponds to optical
transition of $d_{3x^2-r^2}$ and $d_{3y^2-r^2}$ electrons to the
neighboring $Mn^{4+}$ site with parallel spin and therefore related
to the intersite Coulomb interaction. It is separated by a clear
minimum from the higher-lying Mn $3d$ $\rightarrow$ O $2p$
charge-transfer excitations centered at $\sim4$\,eV. On the other
hand, Nd$_{0.55}$Sr$_{0.45}$MnO$_3$ shows metallic-like optical
response. Parallel to the enhanced low-energy spectral weight, the
oscillatory strength of the charge-transfer peak is reduced,
indicating stronger hybridization between Mn $3d$ and O $2p$
electrons, as expected within the framework of double exchange
mechanism.

The evolution of the low-energy optical conductivity spectra with
the variation of the bandwidth is discerned in Fig.~\ref{fig3}.
Three distinct spectral shapes are observed: (i) the two CO/OO
compounds have a charge gap $\Delta_{cg}\approx0.5$\,eV followed by
the broad peak at $E_{peak}=1.4$\,eV; (ii) in case of
Gd$_{0.55}$Sr$_{0.45}$MnO$_3$ with the glassy ground state the
presence of the charge gap is still clear although it is smaller in
magnitude and the $E_{peak}$ is simultaneously reduced; (iii) the
three FM compounds exhibit broad metallic conductivity spectra with
strong incoherent, i.e. only gently $w$-dependent, character up to
$\sim0.5$\,eV which is typical of CMR manganites. Since the spin
sector is fully polarized at low temperatures, orbital fluctuations
are likely responsible for the enhanced scattering amplitude.
Besides the ground state spectra, the optical conductivity just
above $T_c$ is also plotted for the FM compounds. Except for
Nd$_{0.55}$Sr$_{0.45}$MnO$_3$ with the largest bandwidth, they
resemble that of Gd$_{0.55}$Sr$_{0.45}$MnO$_3$ in its glassy ground
state; the charge gap characteristic to the short-range CO/OO phase
is already developed at elevated temperatures above the onset of the
ferromagnetic metallic state. We just note here and discuss later in
detail that the gap of the two CO/OO materials shows a discontinuous
decrease at $T_{CO}$ and that just above the CO/OO transition it is
also close to the ground state value of
Gd$_{0.55}$Sr$_{0.45}$MnO$_3$.

\section{Crossover towards the high-temperature phase}
\label{T-dep}

\begin{figure*}[th!]
\includegraphics[width=4in]{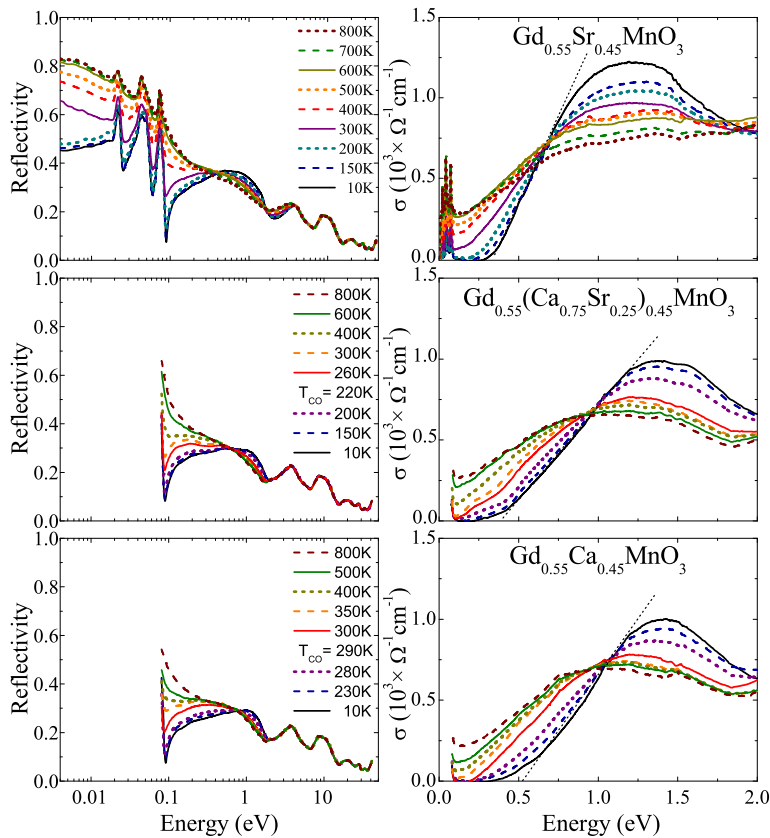}
\caption{(Color online) Reflectivity and conductivity spectra of the
three insulators GCMO, GCSMO, and GSMO at selected temperatures
(left and right panel, respectively). Note the difference in energy
scales in the left (logarithmic) and right (linear) panels. The
linear extrapolation to estimate the gap energy in the ground state
is shown by black dotted lines. The temperature of the CO/OO
transition is indicated in the label.} \label{fig4}
\end{figure*}

Here we turn to the analysis of the temperature-induced
spectral-weight transfer both for the insulating and metallic
compounds. While the charge gap in the CO/OO state of GCMO and GCSMO
shows only a weak temperature dependence up to $T_{CO}$, the loss of
the long-range order results in a sudden decrease of the gap energy
as shown in Fig.~\ref{fig4}. The shape of their optical conductivity
spectra just above the transition resemble that of GSMO in the
ground state. Above $T_{CO}$ the gap is gradually filled by thermal
fluctuations and a fully incoherent low-energy conductivity emerges.
In both compounds the spectral-weight is transfered dramatically
through an equal-absorption (isosbestic) point at
$\omega_{iso}\approx1$\,eV. On the other hand, in GSMO the closing
of the gap occurs in a smooth way. Although in either of the three
insulators the metallic conductivity is not fully recovered at any
temperatures, the development of an incoherent contribution to the
optical conductivity implies the partial liberation of the orbital
degrees of freedom in the $e_g$ sector. More specifically, the
reduced gap in the glassy ground state of GSMO and above $T_{CO}$ in
the CO/OO insulators is due to orbital redistribution of the excess
electrons ($5\%$ relative to the $x=1/2$ commensurate level) which
occupy $d_{3z^2-r^2}$ orbitals of originally $Mn^{4+}$ sites in the
long-range ordered state.\cite{Jirak85} Their optical transitions
dominating the $c$-axis conductivity are characterized by a
remarkably smaller gap energy as observed in
Pr$_{0.6}$Ca$_{0.4}$MnO$_3$.\cite{Okimoto00} The melting of this
CO/OO phase means the emergence of a charge-orbital liquid state in
which the local CO/OO correlations are still dominant and together
with dynamical Jahn-Teller effect preserve electron localization.
Towards high temperatures the electronic state of the three
insulators become similar as manifested in their nearly identical
optical conductivity spectra above $T=600$\,K.

\begin{figure*}[th!]
\includegraphics[width=4in]{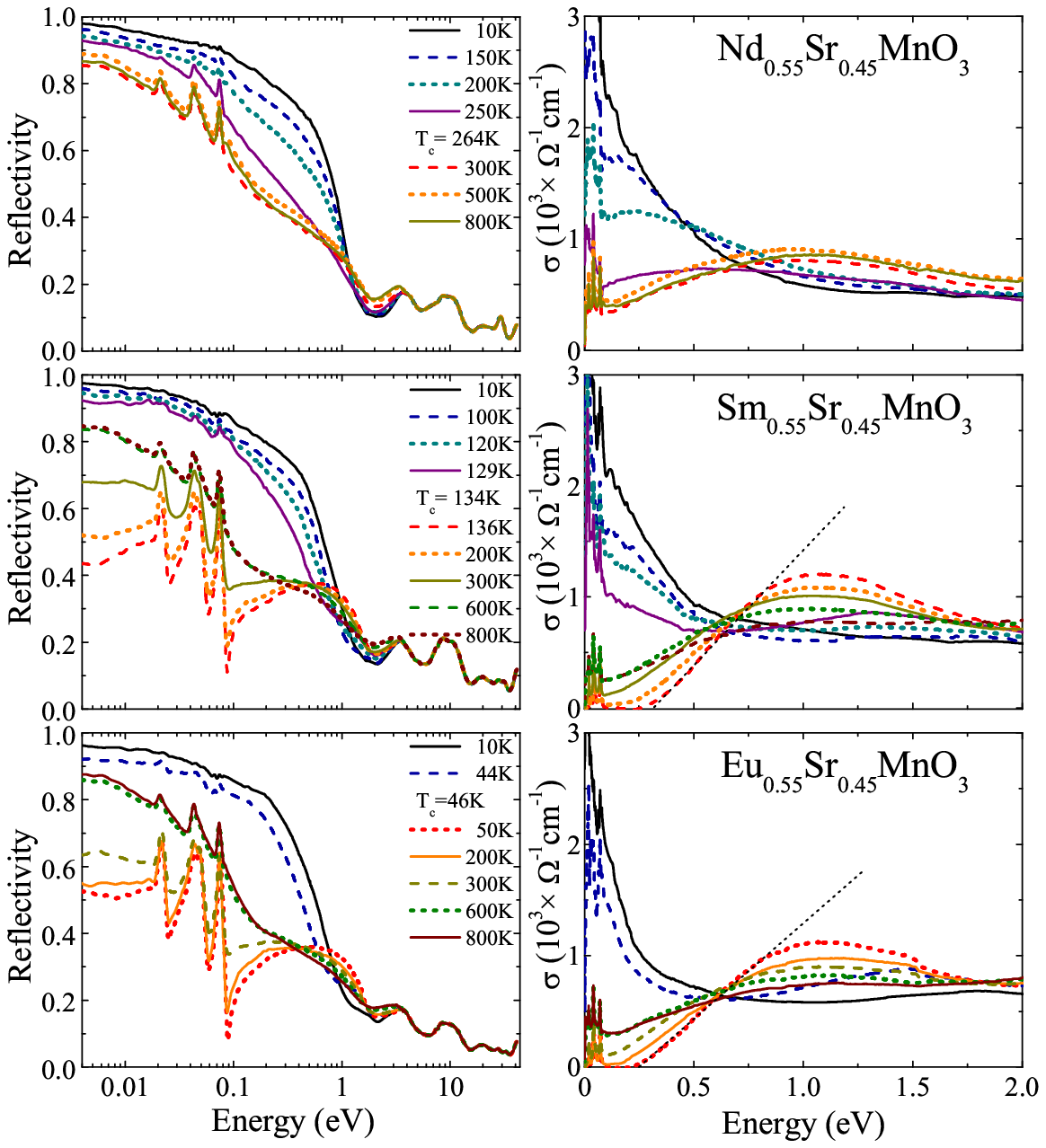}
\caption{(Color online) Reflectivity and conductivity spectra of the
three metals ESMO, SSMO, and NSMO at selected temperatures (left and
right panel, respectively). For ESMO and SSMO, the linear
extrapolation to estimate the gap energy just above the FM
transition is shown by dotted lines.} \label{fig5}
\end{figure*}

At the metallic side of the phase diagram the gradual transfer of
the low-energy spectral weight to the $\sim1$\,eV peak occurs when
$T_c$ is approached (see Fig.~\ref{fig5}). In this case the
isosbestic behavior is also discerned with an equal-absorption point
at $\omega_{iso}\approx0.68$\,eV. This energy scale is considerably
lower than in the CO/OO insulators but close to that of GSMO. The
ferromagnetic to paramagnetic transition is associated with a
dramatic redistribution of the spectral weight, especially in ESMO
and SSMO where the instantaneous opening of the charge gap is
observed above $T_c$. In NSMO with larger bandwidth (and
equivalently with higher Curie temperature) the transition occurs in
a continuous manner and only a pseudo-gap develops in the
paramagnetic (PM) phase. The change in the character of the FM
transition (from first to second order) between SSMO and NSMO can be
well followed in lower panel of Fig.~\ref{phasediag} as well, where
the charge gap is plotted for the whole series of the compounds both
along the $T_{CO}(r,T)$ and $T_c(r,T)$ phase boundaries. With
increasing bandwidth, the gap evaluated at the lowest temperature of
the PM phase shows only tiny variation around $\Delta=0.25$\,eV till
SSMO while it completely vanishes for NSMO. Detailed
magneto-transport and magnetization studies also evidence that the
first-order PM-FM phase boundary is terminated between SSMO and NSMO
and the transition becomes continuous for larger
bandwidth.\cite{Demko08} The presence of first-order phase
boundaries (indicated by thick line in both panels of Fig.~1)
separating the FM state both from the CO/OO and PM phase implies
dramatic change in the orbital character of $e_g$ electrons.
Although it is a central issue in the CMR physics, the orbital
nature of the FM state has not utterly been clarified. In the
present optical study this question is also to address in the
context of the large scattering rate observed in the fully spin
polarized low-temperature state. For the present metals three
possible scenarios have been proposed: (i) a disordered phase
involving both $(x^2-y^2)$ and $(3z^2-r^2)$ orbitals \cite{Tobe04}
or strong ferro-orbital correlations of either (ii) the planar
$(x^2-y^2)$-type \cite{Ishihara97,Khaliullin00} or (iii) the
rod-like $(3z^2-r^2)$-type.\cite{Endoh05} In case of SSMO, the
development of rod-like order is supported by recent neutron
scattering experiments; the anomalously large fourth-neighbor
interaction along $x$ (or $y$ or $z$) direction, $J_4\approx0.6J_1$,
implies the extended one-dimensional exchange path reflecting the
rod-like orbital correlation.\cite{Endoh05} The effect of the
orthorhombic lattice distortion is almost negligible for SSMO (which
is also the case for ESMO and NSMO); the spatial and temporal
fluctuations of the nearly equivalent [001], [010], and [100]
domains prevent the onset of a static long-range orbital ordering
for the investigated temperature region. Therefore, we ascribe the
large scattering rate of the spin-polarized phase to these orbital
fluctuations. The optical isotropy of the $ab$-plane also implies
the lack of long-range order above $T=10$\,K, thus the robust
enhancement of the metallic conductivity in the vicinity of the
ferromagnetic transition is attributable to the nearly
temperature-independent spin-polarization up to $T_c$.

\begin{figure}[th!]
\includegraphics[width=2.4in]{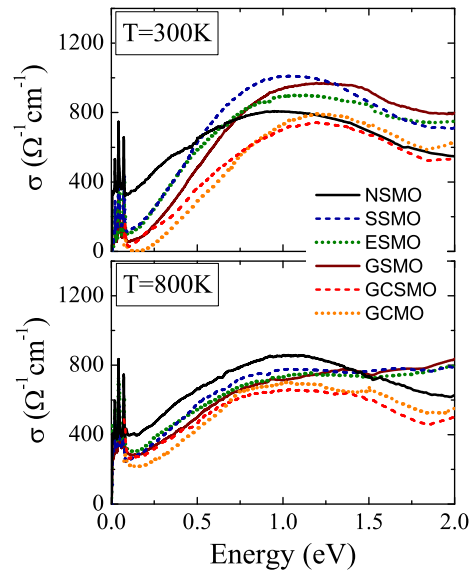}
\caption{(Color online) Comparative plot of the optical conductivity
for the full series of RE$_{0.55}$AE$_{0.45}$MnO$_3$ manganites at
$T=300$\,K and $800$\,K (upper and lower panel, respectively).}
\label{fig6}
\end{figure}

\begin{figure}[th!]
\includegraphics[width=2.35in]{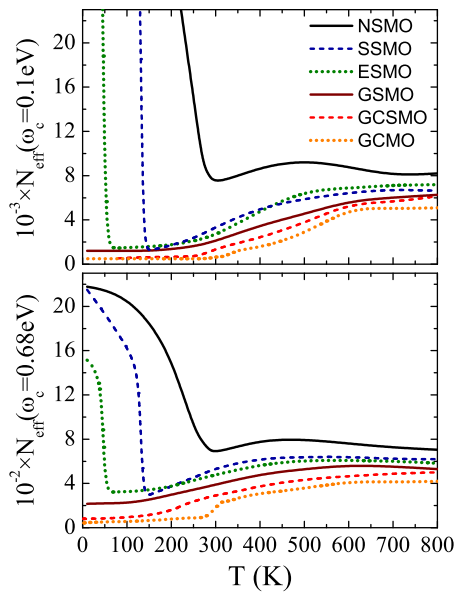}
\caption{(Color online) Temperature dependence of the effective
number of electrons, $N_{eff}(\omega_c)$, at two representative
energies $\omega_c=0.1$\,eV and $0.68$\,eV for the bandwidth-tuned
series of RE$_{0.55}$AE$_{0.45}$MnO$_3$ manganites shown in the
upper and lower panel, respectively.} \label{fig7}
\end{figure}

At room temperature which is slightly above both the FM and CO/OO
phases, the optical conductivity spectra of the respective materials
are close to each other irrespective of the nature of the ground
state. They further converge and become nearly identical at the
highest temperature of the present optical study, $T=800$\,K, as
discerned in Fig.~\ref{fig6}. A systematic analysis of the
low-energy spectral changes is given in Fig.~\ref{fig7} where the
temperature dependence of the spectral weight, i.e. the effective
number of electrons defined as $N_{eff}(\omega_c)=2m_0(\pi
e^2N)^{-1}\int_0^{\omega_c}\sigma(\omega)d\omega$, for
$\omega_c=0.1$\,eV and $0.68$\,eV is simultaneously plotted for each
compound. The former solely represents the metallic conductivity,
while the latter corresponds to energy scale of the equal absorption
point, therefore includes the strong incoherent mid-infrared
contribution. Except for the FM phase where a stronger enhancement
due to the development of coherent conduction is found in
$N_{eff}(\omega_c=0.1eV)$, the temperature dependence of the
spectral weight is rather similar for the both energy scales. This
implies that the low-energy excitations, i.e. the in-gap states, are
fully governed by CO/OO correlations.

The variation of the total spectral weight, represented by
$N_{eff}(\omega_c=0.68eV)$, is also visualized over the whole phase
diagram in the color contour plot of Fig.~\ref{phasediag}. Both the
dramatic increase in $N_{eff}$ below $T_c$ and its reduction below
$T_{CO}$ are clearly manifested in the contour plot. The light-blue
region of the phase diagram characterized by a flat low-energy
conductivity spectrum demonstrates the strong CO/OO fluctuations
extending above all the three underlying phases. The pseudo-gap like
feature of the optical spectra (i.e. the presence of the
mid-infrared peak and the reduced optical weight below it) is
preserved and the metallic conductivity is not recovered up to
$T=800$\,K, indicating the subsisting CO/OO correlations up to
higher temperatures in this critically doped series of CMR
manganites. On the other hand, x-ray experiments on the same
materials could detect the diffuse scattering arising from the
short-range orbital ordering, but no longer than up to
$T=400$\,K.\cite{Tomioka03} This means that above this temperature
the correlation length of the lattice distortion is limited to the
scale of the unit cell. However, infrared optical excitations can
still sensitively pick up the local orbital configuration and thus
give a direct tool for charge-orbital correlation effects in these
materials.

\section{Conclusions}

The aim of the present optical study of nearly half-doped colossal
magnetoresistive manganites RE$_{0.55}$AE$_{0.45}$MnO$_3$ with large
quenched disorder was to characterize the nature of the low-energy
electronic states responsible for the CMR effect. We have
investigated the systematics of the optical conductivity spectrum
over a broad area of the bandwidth vs temperature plane. The ground
state of the compounds ranges from the charge and orbital ordered
insulator through the spin glass to the ferromagnetic metal. The key
role of critically enhanced phase fluctuations, most typically the
short-range CO/OO correlation, was found in the paramagnetic region
of the phase diagram above all the ground-state phases. This
charge-orbital liquid state is characterized by a pseudo-gap below
$E_{peak}\approx1$\,eV in the optical conductivity and flat
low-energy optical conductivity spectrum over a large variation of
the bandwidth. It is robust against thermal agitations and extends
up to as high temperature as $T=800$\,K. The onset of the long-range
charge order is accompanied by an instantaneous increase of the gap.
The emergence of the ferromagnetic state is manifested in the
optical spectra as a first order insulator to metal transition for
compounds with moderate bandwidth while it becomes a continuous
transition on the largest bandwidth side. Unusually large scattering
rate of the metallic carriers is observed in the ferromagnetic state
which is attributed to orbital correlation probably with rod-like
character.

\section*{Acknowledgement}

The authors are grateful to N. Nagaosa and S. Onoda for enlightening
discussions. This work was supported in part by a Graint-In-Aid for
Scientific Research, MEXT of Japan. I. K. acknowledges support from
Bolyai J\'anos Fellowship and the Hungarian Scientific Research
Funds OTKA under grant Nos. F61413 and K62441.

%
%

\end{document}